\documentclass[final,authoryear,1p,times]{elsarticle}
\usepackage{graphics}
\usepackage{color}
\usepackage{amsmath}
\usepackage{amssymb}
\usepackage{psfrag}

\def\th{\vec{\theta}} 
\def\u{\vec{U}}
\def\x{\vec{x}}

\def\u{{\bf U}} 

\def\V2{V_2}
\def\V2ij{V_{2ij}}

\def\V{\mathcal{V}}

\def\lsim{~\rlap{$<$}{\lower 1.0ex\hbox{$\sim$}}}

\def\gsim{~\rlap{$>$}{\lower 1.0ex\hbox{$\sim$}}}

\journal{New Astronomy}

\begin{document}

\begin{frontmatter}

\title{Validating a novel angular power spectrum estimator using simulated low frequency radio-interferometric data {\footnote{$\copyright$2017. This manuscript version is made available under the CC-BY-NC-ND 4.0 license http://creativecommons.org/licenses/by-nc-nd/4.0/}}}

\author[]{Samir Choudhuri \corref{cor1}}
\ead{samir11@phy.iitkgp.ernet.in}
\address{Department of Physics,  \& Centre for Theoretical Studies, IIT Kharagpur,  Pin: 721302, India\\
National Centre For Radio Astrophysics, Post Bag 3, Ganeshkhind, Pune 411007, India} 

\author[]{Nirupam Roy}
\address{Department of Physics, Indian Institute of Science, Bangalore 560012, India} 

\author[]{Somnath Bharadwaj}
\address{Department of Physics and Meteorology \& Centre for Theoretical Studies, IIT Kharagpur, 721302  India} 

\author[]{Sk. Saiyad Ali}
\address{Department of Physics,Jadavpur University, Kolkata 700032, India}   

\author[]{Abhik Ghosh}
\address{Dept of Physics and Astronomy, University of the Western Cape, Robert Sobukwe Road, Bellville 7535, South Africa\\
SKA SA, The Park, Park Road, Pinelands 7405, South Africa} 

\author[]{Prasun Dutta}
\address{Department of Physics, IIT (BHU), Varanasi 221005, India}

\begin{abstract}
The ``Tapered Gridded Estimator'' (TGE) is a novel way to directly estimate the 
angular power spectrum from radio-interferometric visibility data that reduces 
the computation by efficiently gridding the data, consistently removes the 
noise bias, and suppresses the foreground contamination to a large extent by 
tapering the primary beam response through an appropriate convolution in the 
visibility domain. Here we demonstrate the effectiveness of TGE in recovering 
the diffuse emission power spectrum through numerical simulations. We present 
details of the simulation used to generate low frequency visibility data for 
sky model with extragalactic compact radio sources and diffuse Galactic synchrotron emission. We then use different imaging strategies to identify the most effective option of point source subtraction and to study the underlying diffuse emission. Finally, we 
apply TGE to the residual data to measure the angular power spectrum, and 
assess the impact of incomplete point source subtraction in recovering the 
input power spectrum $C_{\ell}$ of the synchrotron emission. This estimator is 
found to successfully recovers the $C_{\ell}$ of input model from the residual 
visibility data. These results are relevant for measuring the diffuse emission like 
the Galactic synchrotron emission. It is also an important step towards 
characterizing and removing both diffuse and compact foreground emission in 
order to detect the redshifted $21\, {\rm cm}$ signal from the Epoch of 
Reionization.
\end{abstract} 


\end{frontmatter}

\section{Introduction}

A detailed investigation and analysis of the Galactic diffuse synchrotron 
emission power spectrum can be used to study the distribution of cosmic ray 
electrons and the magnetic fields in the interstellar medium (ISM) of the Milky Way, and is very 
interesting in its own right \citep{Waelkens,Lazarian,iacobelli13}. On the 
other hand, at a very different scale, observations of redshifted $21\, {\rm 
cm}$ radiation from neutral hydrogen (HI) hold the potential of tracing the 
large scale structure of the Universe over a large redshift range of $200 \ge z \ge 
0$. Accurate cosmological HI tomography and power spectrum measurement, 
particularly from the Epoch of Reionization (EoR), by ongoing or future 
low-frequency experiments will provide us a significant amount of information 
about various astrophysical and cosmological phenomena to enhance our present 
understanding of the Universe. Interestingly, since one of the main challenges 
in statistical detection of the redshifted $21\, {\rm cm}$ signal arises from 
the contamination by Galactic and extragalactic ``foregrounds'' \citep{shaver99,dmat1,santos05}, these two aspects are also quite related. The two major 
foreground components for cosmological HI studies are (1) the bright compact 
(``point'') sources and (2) the diffuse Galactic synchrotron emission 
\citep{ali08,paciga11,bernardi09,ghosh150,iacobelli13}. Detection of the weak 
cosmological HI signal will require a proper characterization and removal of point 
sources as well as this diffuse foregrounds. 

Naturally, a significant amount of effort has gone into addressing the problem 
of foreground removal for detecting the $21\, {\rm cm}$ power spectrum from EoR \citep{morales06,jelic08,liu09a,liu09,harker10,mao12,liu2,chapman12,paciga13}. In contrast, foreground 
avoidance \citep{datta10a,vedantham12,morales12,trott1,parsons12,pober13,dillon13,hazelton13,thyag13,liu14a,liu14b,zali15,trott16} 
is an alternative approach based on the idea that contamination from any 
foreground with smooth spectral behaviour is confined only to a wedge in 
cylindrical $(k_{\perp}, k_{\parallel})$ space due to chromatic coupling of an 
interferometer with the foregrounds. The HI power spectrum can be estimated 
from the uncontaminated modes outside the wedge region termed as the $EoR \,\, 
window$ where the HI signal is dominant over the foregrounds. With their merits 
and demerits, these two approaches are considered complementary \citep{chapman16}. 

Here we have considered the issue of estimating the angular power spectrum 
directly form the radio-interferometric ``visibility'' data. In this endeavor, 
we have developed a novel and fast estimator of angular power spectrum that 
consistently avoids the noise bias, and tested it with simulated diffuse 
Galactic synchrotron emission \citep{samir14}. Here, we have further developed 
the simulations to include the point sources in the sky model (as well as 
instrumental noise) to investigate the effectiveness of the estimator of recovering the diffuse 
emission power spectrum in presence of the point sources. This paper describes 
the details of the simulations and analysis, including the adopted point source 
modeling and subtraction strategies, and their effects on the residual diffuse 
emission. We demonstrate that, by using this newly developed Tapered Gridded 
Estimator (hereafter TGE), we can avoid some of the complications of wide field 
low frequency imaging by suitably tapering the primary beam during power 
spectrum estimation. A companion paper has reported the usefulness of the new 
estimator in recovering the diffuse emission power spectrum from the residual 
data in such situation \citep{samir16a}. A further generalization of the 
estimator to deal with spherical and cylindrical power spectrum is presented in 
\citet{samir16b}. Please note that this is part of a coherent effort of 
end-to-end simulation of realistic EoR signal and foreground components, and 
finally using suitable power spectrum estimator to recover the signal. However, 
even though these exercises are in the context of EoR experiments, for the sake 
of simplicity, we have so far not included the weak cosmological signal in the 
model. Here we establish the ability of the developed estimator to recover the 
diffuse emission power spectrum accurately after point source subtraction. 
Thus, apart from EoR experiments, these results are also relevant in more 
general situation, e.g. detailed study of Galactic synchrotron emission 
\citep{samir16c}.

The current paper is organized as follows. In Section 2, we discuss the details 
of the point source and diffuse emission simulation. Section 3 and 4 present 
the analysis using different CLEANing options for point source subtraction and 
the results of power spectrum estimation. Finally, we present summary and 
conclusions in section 5.

\section{Multi-frequency Foreground Simulation}
\label{sec:simu}

In this section we describe the details of the foreground simulation to produce the sky model for generating visibilities for low radio frequency observation 
with an interferometer. Even if the simulation, described in this paper, is 
carried out specifically for $150 \ {\rm MHz}$ observation with the Giant 
Metrewave Radio Telescope (GMRT), it is generic and can easily be extended to 
other frequency and other similar telescopes including the Square Kilometre 
Array (SKA). 

Earlier studies \citep{ali08,paciga11} have found that, for $150\, {\rm MHz}$ 
GMRT small field observations, the bright compact sources are the dominating 
foreground component for EoR signal at the angular scales $\le  4^{\circ}$, the 
other major component being the Galactic diffuse synchrotron emission 
\citep{bernardi09,ghosh150,iacobelli13}. We build our foreground sky model 
keeping close to the existing observational findings. The sky model includes 
the main two foreground components (i) discrete radio point sources and (ii) 
diffuse Galactic synchrotron emissions. The contributions from these two 
foregrounds dominate in low frequency radio observations and their strength is 
$\sim 4-5$ orders of magnitude larger than the $\sim 20-30\, {\rm mK}$ 
cosmological $21$-cm signal \citep{ali08,ghosh150}. Galactic and extragalactic 
free-free diffuse emissions are also not included in the model, though each of 
these is individually larger than the HI signal.

\subsection{Radio Point Sources}
\label{sec:ptsrc}

\begin{figure*}
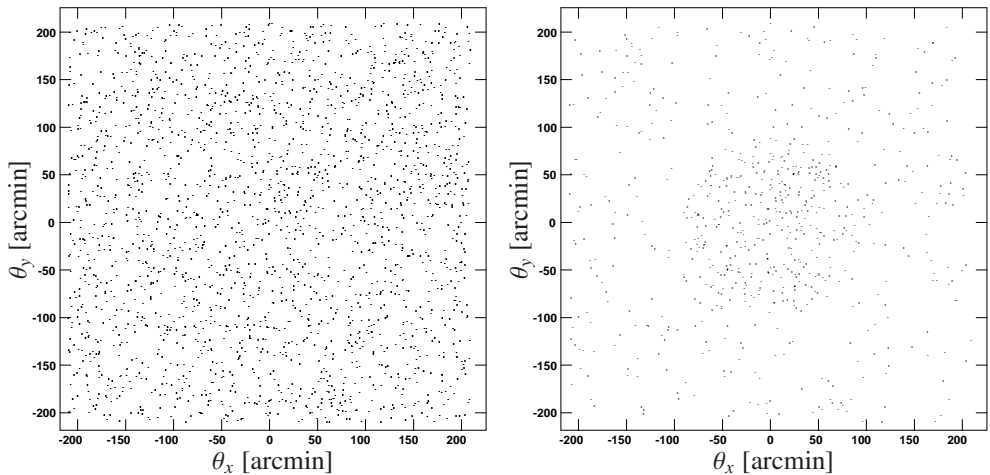

\begin{center}
\psfrag{thetax}[t][c][1][0]{$\theta_x$ $[{\rm arcmin}]$}
\psfrag{thetay}[c][c][1][0]{$\theta_y$ $[{\rm arcmin}]$}
\includegraphics[width=65mm,angle=0]{ptsrc1small1.ps}
\includegraphics[width=65mm,angle=0]{ptsrc696.ps}
\caption{The angular position of the simulated point sources over a $7^{\circ} \times 7^{\circ}$ region. The left panel shows positions of all $2215$ sources over this whole field, and the right panel shows $696$ sources after applying a flux density cutoff. The number of point sources in the flux density range $9\,{\rm mJy}$ to $1\,{\rm Jy}$ inside the FWHM of the primary beam is $N_{in}=353$ and outside of the FWHM with flux density more than 100 {\rm mJy} is $N_{out}=343$.}
\label{fig:ptsrc}
\end{center}
\end{figure*}

Most of the earlier exercise of numerical simulation conducted so far have not 
included the bright point source foreground component in the multi-frequency 
model. In such analysis, it is generally assumed that the brightest point 
sources are perfectly subtracted from the data before the main analysis, and 
the simulated data contains only faint point sources and other diffuse 
foreground components, HI signal and noise. We, however, simulate the point 
source distribution for sky model using the following differential source 
counts obtained from the GMRT $150 \ {\rm MHz}$ observation \citep{ghosh150}:
\begin{equation}
\frac{dN}{dS} = \frac{10^{3.75}}{\rm Jy.Sr}\,\left(\frac{S}{\rm Jy}\right)^{-1.6} \,.
\label{eq:dnds}
\end{equation}
The full width half maxima (FWHM) of the GMRT primary beam (PB) at $150 \ {\rm 
MHz}$ is $\approx3.1^{\circ}$. To understand and quantify how the bright point 
sources outside the FWHM of the PB affect our results, we consider here a 
larger region ($7^{\circ} \times 7^{\circ}$) for point source simulation. 
Initially, $2215$ simulated point sources, with flux density in the range 
$9\,{\rm mJy}$ to $1\,{\rm Jy}$ following the above mentioned source count, are 
randomly distributed over this larger region. Out of those sources, $353$ are 
within ${95}^{'}$ from the phase centre (where the PB response falls by a 
factor of $e$). We note that the antenna response falls sharply after this 
radius. For example, the primary beam response is $\lesssim$ 0.01 in the first 
sidelobe. Hence, outside this ``inner'' region, only sources with flux density 
greater than $100 \, {\rm mJy}$ are retained for the next step of the 
simulation. In the outer region, any source fainter than this will be below the 
threshold of point source subtraction due to primary beam attenuation. With 
$343$ sources from the ``outer'' region, we finally include total $696$ sources 
in our simulation. Figure~\ref{fig:ptsrc} shows the angular positions of all 
$2215$ sources over this region, as well as of the $696$ sources after the flux 
density restriction. Note that, we have assumed all the sources are unresolved 
at the angular resolution of our simulation. In reality, there will also be 
extended sources in the field. Some of the extended sources can be modelled 
reasonably well as collection of multiple unresolved sources. However, other 
complex structures will need more careful modelling or masking, and are not 
included in this simulation for simplicity.

The flux density of point sources changes across the frequency band of 
observation. We scale the flux density of the sources at different frequencies 
using the following relation,
\begin{equation}
S_{\nu}=S_{\nu_0}\left(\frac{\nu}{\nu_0}\right)^{-\alpha_{\rm ps}} \,
\label{eq:fluxscale}
\end{equation}
where $\nu_0=150\,{\rm MHz}$ is the central frequency of the band, $\nu$ 
changes across the bandwidth of $16{\rm MHz}$ and $\alpha_{\rm ps}$ is the 
spectral index of point sources. The point sources are allocated a randomly 
selected spectral index uniform in the range of $0.7$ to $0.8$ 
\citep{jackson05,randall12}. Please note that the subsequent point source 
modeling and subtraction are carried out in such a way that the final outcomes 
do not depend on the exact distribution function of the spectral index.

\subsection{Diffuse Synchrotron Emission}
\label{sec:diff}

In this section, we first describe the simulation of the diffuse Galactic 
synchrotron emission which are used to generate the visibilities. The angular 
slope $\beta$ of the angular power spectrum of diffuse Galactic synchrotron 
emission is within the range $1.5$ to $3$ as found by all the previous 
measurements at frequency range $0.15 -94 \, {\rm GHz}$ (e.g. 
\citealt{tegmark96,tegmark2000,giardino02,bennett03,laporta,bernardi09,ghosh150,iacobelli13,samir16c}). For the purpose of this paper, we assume that the fluctuations 
in the diffuse Galactic synchrotron radiation are coming from a statistically 
homogeneous and isotropic Gaussian random field whose statistical properties 
are completely specified by the angular power spectrum. We construct our sky 
model of the diffuse Galactic synchrotron emission using the measured angular power 
spectrum at $150\,{\rm MHz}$ \citep{ghosh150}
\begin{equation}
C^M_{\ell}(\nu)=A_{\rm 150}\times\left(\frac{1000}{\ell} \right)^{\beta}\times\left(\frac{\nu}{150{\rm MHz}}\right)^{-2\alpha_{\rm syn}}   \,,
\label{eq:cl150}
\end{equation}
where $\nu$ is the frequency in ${\rm MHz}$, $A_{\rm 150}=513 \, {\rm mK}^2$ 
and $\beta=2.34 $ adopted from \citet{ghosh150} and $\alpha_{\rm syn}=2.8$ 
from \citet{platania1}. The diffuse emissions are generated in a $1024 \times 
1024$ grid with angular grid size of $\sim 0.5^{'}$, covering a region of 
$\ 8.7^{\circ} \times 8.7^{\circ}$. This axis dimension is $\approx 2.8$ times 
larger than the FWHM of the GMRT primary beam.

\begin{figure*}
\begin{center}
\psfrag{thetax}[c][c][0.6][0]{$\theta_x$ $[{\rm arcmin}]$}
\psfrag{thetay}[c][c][0.6][0]{$\theta_y$ $[{\rm arcmin}]$}
\psfrag{cl}[b][t][1][0]{$C_{\ell} [mK^2]$}
\psfrag{U}[c][c][1][0]{$\ell$}
\psfrag{model}[c][l][1][0]{Model}
\psfrag{beam}[r][r][1][0]{with PB}
\psfrag{nobeam}[r][r][1][0]{without PB}
\includegraphics[height=70mm,angle=0]{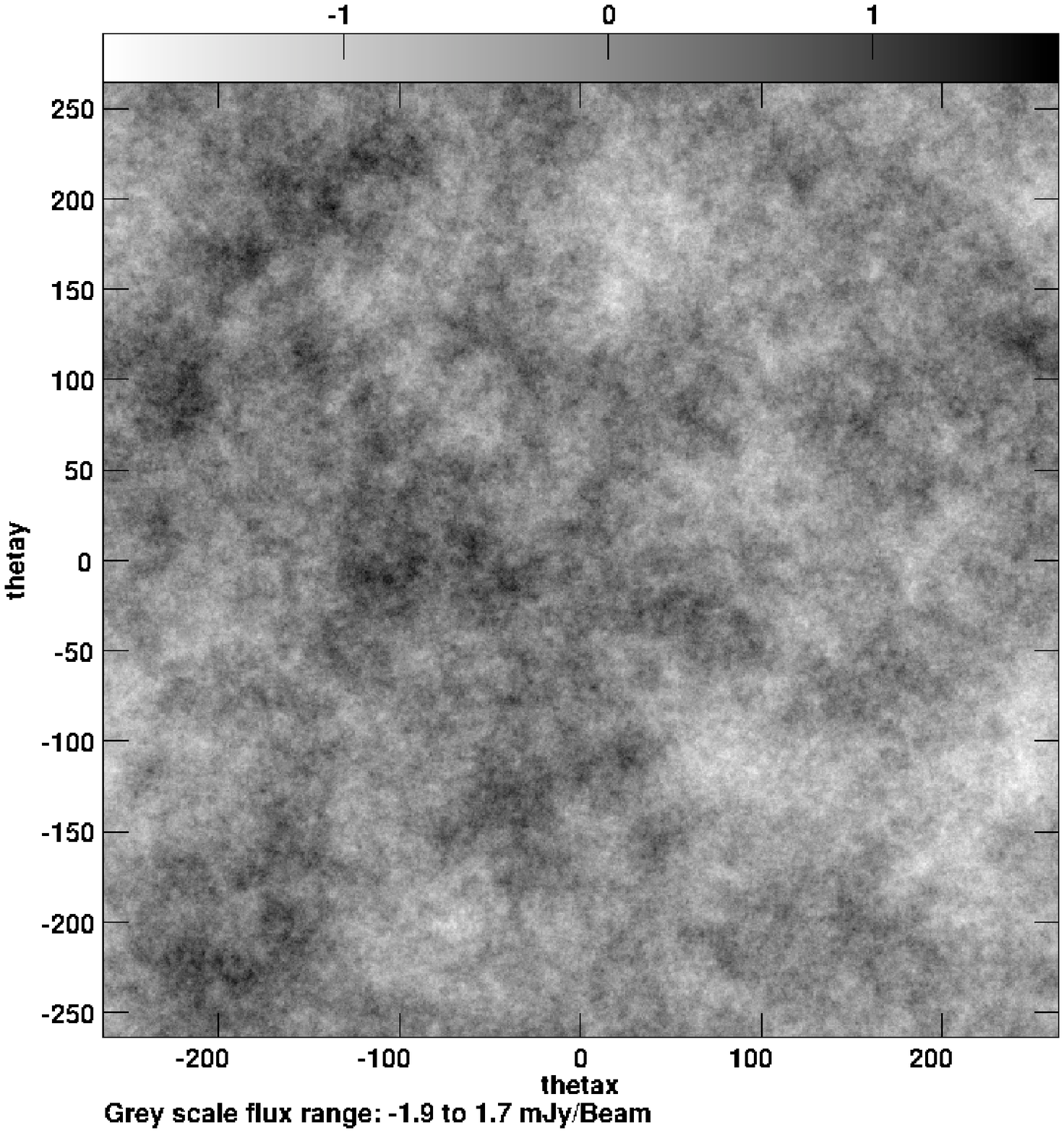}
\includegraphics[height=70mm,angle=0]{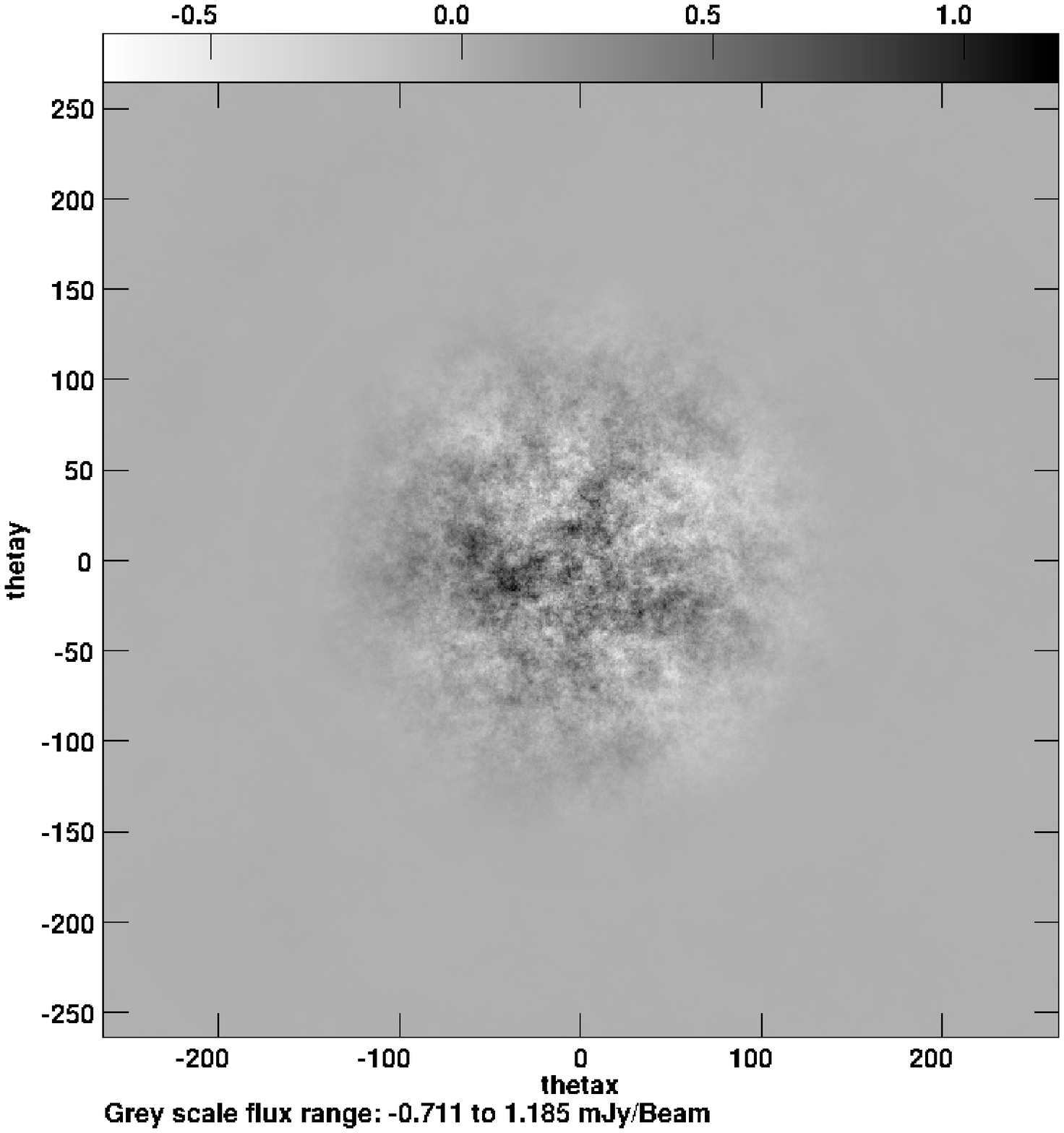}
\caption{The simulated $150 \, {\rm MHz}$ diffuse emission map before (left panel) and after (right panel) multiplying the GMRT primary beam. The images are $8.7^{\circ} \times 8.7^{\circ}$ with a grid size $\sim 0.5^{'}$, and the grey scale is in ${\rm mJy/Beam}$.}
\label{fig:diff}
\end{center}
\end{figure*}

To simulate the diffuse emission, we mainly followed the same procedure as 
discussed in \citet{samir14}. We first create the Fourier components of the 
temperature fluctuations on a grid using
\begin{equation}
\Delta \tilde{T}(\u,\nu_0)=\sqrt{\frac{\Omega \, C^M_{\ell}(\nu_0)}{2}}[x(\u)+iy(\u)],
\label{eq:ran}
\end{equation}
where $\Omega$ is the total solid angle of the simulated area, and $x(\u)$ and 
$y(\u)$ are independent Gaussian random variables with zero mean and unit 
variance. Then, we use the Fastest Fourier Transform in the West (hereafter 
FFTW) algorithm \citep{frigo05} to convert $\Delta\tilde{T}(\u,\nu_0)$ to 
the brightness temperature fluctuations $\delta T(\th,\nu_0)$ or, equivalently, 
the intensity fluctuations $\delta I(\th,\nu_0)$ on the grid. The intensity 
fluctuations $\delta I(\th,\nu) = (2 k_B/\lambda^2)\, \delta T(\th,\nu) $ can 
be calculated using the Raleigh-Jeans approximation which is valid at the 
frequency of our interest.

Finally, we generate the specific intensity fluctuations at any other frequency 
within the observing band from that of the reference frequency using the 
scaling relation 
\begin{equation}
\delta I(\th,\nu)= \left(2 k_B/\lambda^2\right)\, \delta T(\th,\nu_0)\left(\frac{\nu}{\nu_0}\right)^{-\alpha_{\rm syn}} \,.
\label{eq:tempscale}
\end{equation}
In general, the spectral index $\alpha_{\rm syn}$ of the diffuse emission may 
have a spatial variation and the synchrotron power spectrum may be different at 
different frequencies. However, the effect of this on point source subtraction 
is expected to be negligible, and the final results do not depend on the 
constancy of the synchrotron power spectrum slope. Here, we assume that the 
value of $\alpha_{\rm syn}$ is fixed over the whole region and across the 
observation band in the multi-frequency simulation. 

\subsection{GMRT Primary beam}
We model the PB of GMRT assuming that the 
telescope has an uniformly illuminated circular aperture of $45\,{\rm m}$ 
diameter (D) whereby the primary beam pattern is given by,
\begin{equation}
{\cal A}(\th,\,\nu) = 
\left[ \left(\frac{2 \lambda}{\pi\theta D} \right)
J_1\left(\frac{\pi\theta D}{\lambda}\right) \right]^2 
\label{eq:b1} 
\end{equation}
where $J_1$ is the Bessel function of the first kind of order one. The primary 
beam pattern is normalized to unity at the pointing center $[{\cal A}(0)=1]$. The central part of the model PB (eq.~\ref{eq:b1}) is a reasonably good approximation to the actual PB of the GMRT antenna, whereby, it may vary at the outer region. In our analysis, we taper the outer region through a window function, hence the results are not significantly affected by the use of this approximate model PB.

Figure~\ref{fig:diff} shows one realization of the 
intensity fluctuations $\delta I(\th,\nu_0)$ map at the central frequency 
$\nu_0=150\,{\rm MHz}$ with and without multiplication of the GMRT primary 
beam. The PB only affect the estimated angular power spectrum at large angular 
scales ($\lesssim 45~\lambda$) which is shown in Figure 3 of \citet{samir14}. Using a large number of realizations of the diffuse 
emission map, we find that the recovered angular power spectrum is in good 
agreement with the input model power spectrum (eq. \ref{eq:cl150}) at the 
scales of our interest $(\ell\sim300-2\times10^4)$.

\subsection{Simulated GMRT Observation}

\begin{figure}
\begin{center}
\hspace*{-1cm}
\includegraphics[width=60mm,angle=-90]{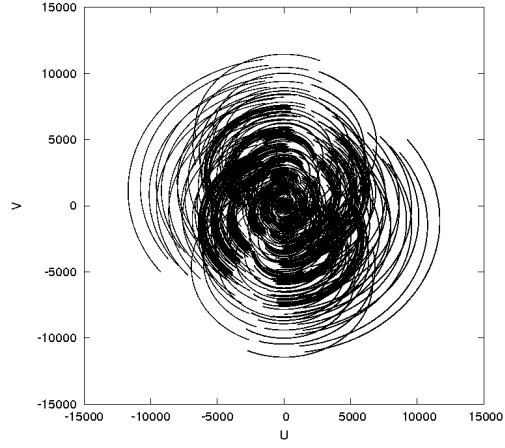}
\caption{The GMRT $uv$ coverage with phase centre at R.A.=$10{\rm h} 46{\rm m} 00{\rm s}$ Dec=$59^{\circ} 00^{'} 59^{''}$ for a total observation time of $8{\rm hr}$. Note that $u$ and $v$ are projected antenna separations in units of wavelength at the central frequency $150\, {\rm MHz}$.}
\label{fig:uvtrack}
\end{center}
\end{figure}

The simulations are generated keeping realistic GMRT specifications in mind, 
though these parameters are quite general, and similar mock data for any other 
telescope can be generated easily. The GMRT has $30$ antennas. The diameter of 
each  antenna is $45\rm m$. The projected shortest baseline at the GMRT can be  
$60\rm m$, and the longest baseline is $\rm 26 \,km$. The instantaneous 
bandwidth is $16 \, {\rm MHz}$, divided into $128$ channels, centered at $150 
\, {\rm MHz}$. We consider all antennas pointed to an arbitrary field located 
at R.A.=$10{\rm h} 46{\rm m} 00{\rm s}$ Dec=$59^{\circ} 00^{'} 59^{''}$ for a 
total of $8 \, {\rm hr}$ observation. The visibility integration time was 
chosen as $16 \,{\rm s}$. The mock observation produces $783000$ samples per 
channels in the whole $uv$ range. Figure~\ref{fig:uvtrack} shows the full $uv$ coverage at central frequency for 
the simulated GMRT Observation. 

The angular power spectrum of the diffuse synchrotron emission (eq. 
\ref{eq:cl150}) declines with increasing baseline $U=\mid \u \mid$ (where 
$\ell = 2 \pi U$), and drops significantly at the available longest baseline. 
Hence, for our simulation, the contributions of the diffuse emission have been 
taken from only baselines $U \le 3,000~\lambda$ to reduce the computation time. 
To calculate the visibilities, we multiply the simulated intensity fluctuations 
$\delta I(\th, \,\nu)$ with the PB (eq.~\ref{eq:b1}), and we use 2-D FFTW of 
the product in a grid. For each sampled baseline $U \le 3,000~\lambda$, we 
interpolate the gridded visibilities to the nearest baseline of the $uv$ track 
in Figure~\ref{fig:uvtrack}. We notice that the $w$-term does not have 
significant impact on the estimated angular power spectrum of diffuse 
synchrotron emission \citep{samir14}. But, to make the image properly and also 
to reduce the sidelobes of the point spread function (or the synthesized beam), 
it is necessary to retain the $w$-term information. The $w$-term also improves 
the dynamic range of the image and enhances the precision of point source 
subtraction. We use the full baseline range to calculate the contribution from 
the point sources. The sky model for the point sources is multiplied with PB 
${\cal A}(\th,\,\nu)$ before calculating the visibilities. Using the small 
field of view approximation, the visibilities for point sources are computed at 
each baseline by incorporating the $w$ term:
\begin{equation}
V(\u,\nu) \approx \int d^2 \theta  {\cal A}(\th,\,\nu) \,  \delta I(\th,\nu) \, e^{- 2 \pi i\big(u\theta_x+v\theta_y+w\big(\sqrt{1-\theta_x^2-\theta_y^2}-1\big)\big)} \, .
\label{eq:vis}
\end{equation}
The system noise of the interferometer is considered to be independent at 
different baselines and channels, and is modelled as Gaussian random variable. 
We add independent Gaussian random noise to both the real and imaginary parts 
of each visibility. For a single polarization, the theoretical rms noise in the 
real or imaginary part of a measured visibility is 
\begin{equation}
\sigma
=\frac{\sqrt2k_BT_{sys}}{A_{eff}\sqrt{\Delta \nu
      \Delta t}}
\label{eq:rms}
\end{equation}
where $T_{sys}$ is the total system temperature, $k_B$ is the Boltzmann 
constant, $A_{eff}$ is the effective collecting area of each antenna, $\Delta 
\nu$ is the channel width and $\Delta t$ is correlator integration time \citep{thompson}. For $\Delta\nu=125\,{\rm kHz}$ and $\Delta t = 16\,{\rm sec}$, the 
rms noise comes out to be $\sigma_n=1.03 \, {\rm Jy}$ per single polarization 
visibility for GMRT. The two polarizations are assumed to have identical sky 
signals but independent noise contribution.

In summary, the simulated visibilities for the GMRT observation are sum of two 
independent components namely the sky signal and the system noise. As outlined 
above, the realistic sky signal includes the extragalactic point sources and 
the Galactic diffuse synchrotron emission. The visibility data does not contain 
any calibration errors, ionospheric effects and radio-frequency interference 
(RFI), and a detailed investigation of these effects are left for future work.

\section{Data Analysis}
\label{sec:data}

Our next goal is to analyze the simulated data described above to recover the 
statistical properties of the diffuse emission, and compare those with the 
known input model parameters. As mentioned earlier, to estimate the power 
spectrum of the diffuse emission, our approach is to first remove the point 
source foreground accurately. This requires imaging and deconvolution to model 
the point sources, and then subtracting them from the data. In reality, there 
are many issues which make an accurate subtraction of point sources from radio 
interferometric wide-field synthesis images challenging. These include residual 
gain calibration errors \citep{datta10}, direction dependence of the 
calibration due to instrumental or ionospheric/atmospheric conditions 
\citep{intema09,yata12}, the effect of spectral index of the sources 
\citep{rau11}, frequency dependence and asymmetry of the primary beam response, 
varying point spread function (synthesized beam) of the telescope 
\citep{liu09a,morales12,ghosh150}, high computational expenses of imaging a 
large field of view, and CLEANing a large number of point sources 
\citep[particularly severe at low radio frequency images,][]{pindor11} etc. 
Note that these issues are more prominent at low radio frequencies due to a 
comparatively large field of view as well as a large number of strong point 
sources and bright Galactic synchrotron emission. Hence, foreground is one of 
the major problem particularly in the context of EoR and post-EoR cosmological 
HI studies with the current and future telescopes (e.g. GMRT{\footnote{Giant Metrewave Radio Telescope; http://www.gmrt.ncra.tifr.res.in}}, LOFAR{\footnote{Low Frequency Array; http://www.lofar.org}}, MWA{\footnote{Murchison Wide-field Array; http://www.mwatelescope.org}}, PAPER{\footnote{Precision Array to Probe the Epoch of Reionization; http://astro.berkeley.edu/dbacker/eor}}, PaST{\footnote{Primeval Structure Telescope; http://web.phys.cmu.edu/~past}}, HERA{\footnote{Hydrogen Epoch of Reionization Array; http://reionization.org/}}, and SKA{\footnote{Square Kilometer Array; http://www.skatelescope.org}}).

Earlier, \citet{datta09, datta10} have studied the effect of calibration errors 
in bright point source subtraction. They have concluded that, to detect the 
EoR signal, sources brighter than $1\, {\rm Jy}$ should be subtracted with a 
positional accuracy better than 0.1 arcsec if calibration errors remain 
correlated for a minimum time $\sim$ 6 hours of observation. On the other hand, 
\citet{bowman09} and \citet{liu09} have reported that point sources should be 
subtracted down to a $10-100\,{\rm mJy}$ threshold in order to detect the $21\, 
{\rm cm}$ signal from the EoR. It has also been recently demonstrated using 
both simulated and observed data from MWA that foreground (particularly the 
point sources) must be considered as a wide-field contaminant to measure the 
$21\, {\rm cm}$ power spectrum \citep{pober16}. The polarized galactic 
synchrotron emission is expected to be Faraday-rotated along the path, and it 
may acquire additional spectral structure through polarization leakage at the 
telescope. This is a potential complication for detecting the HI signal 
\citep{jelic10,moore13}. To cope with the capabilities of current and 
forthcoming radio telescopes, recently there have been a significant progress 
in developing calibration, imaging and deconvolution algorithms 
\citep{bhatnagar13,corn08} which can now handle some of the above-mentioned 
complications.

Keeping aside calibration errors, the problem of subtracting point sources 
ultimately reduces to a problem of deconvolution of point sources, in presence 
of diffuse (foreground and/or cosmological HI signal) emission, to fit their 
position, flux density and spectral property as accurately as the instrumental 
noise permits. The optimum strategy of modeling and subtracting point sources 
in presence of diffuse emission is an open question in the general context of 
interferometric radio frequency data analysis. In this paper, we take up a 
systematic analysis of the $150\,{\rm MHz}$ simulated data to quantify effect 
of incomplete spectral modeling and of different deconvolution strategies to 
model and subtract point sources for recovering the diffuse emission power 
spectrum. In particular, we demonstrate the advantage of the power spectrum 
estimator that we have used (TGE) which allow us to avoid wide field imaging in 
order to subtract the point sources from the outer part of the field of view. 
As a result, it also takes care of, at least to a large extent, issues like 
asymmetry of the primary beam, direction dependence of the calibration for the 
outer region of the field of view and high computational expenses of imaging 
and removing point sources from a large field of view etc. Below we describe 
the details of the imaging and point source subtraction used to produce the 
residual visibility data for power spectrum estimation.

\subsection{Imaging and Power spectrum Estimation}

For our analysis, we use the Common Astronomy Software Applications (CASA) 
\footnote{http://casa.nrao.edu/} to produce the sky images from the simulated 
visibility data. To make a CLEAN intensity image, we use the Cotton-Schwab 
CLEANing algorithm \citep{schwab84} with Briggs weighting and robust parameter 
0.5, and with different CLEANing thresholds and CLEANing boxes around point 
sources. The CLEANing is also done with or without multifrequency synthesis 
(MFS; \citealt{sault94,conway90,rau11}). If MFS is used during deconvolution, 
it takes into account the spectral variation of the point sources using Taylor 
series coefficients as spectral basis functions. In a recent paper 
\citet{offringa16} suggest that CASA's MS-MFS algorithm can be used for better 
spectral modelling of the point sources. The large field of view 
($\theta_{FWHM}= 3.1^{\circ}$) of the GMRT at $150\, {\rm MHz}$ lead to 
significant amount of errors if the non-planar nature of the GMRT antenna 
distribution is not taken into account. For this purpose we use $w-$projection 
algorithm \citep{corn08} implemented in CLEAN task within the CASA. For 
different CLEANing strategies, we assess the impact of point sources removal in 
recovering the input angular power spectrum $C_{\ell}$ of diffuse Galactic 
synchrotron emission from residual $uv$ data. Effectively, by CLEANing with 
these different options, we identify the optimum approach to produce the best 
model for point source subtraction and $C_{\ell}$ estimation. We investigate 
the CLEANing effects in the image domain by directly inspecting the ``residual 
images'' after the point source subtraction, and also in the Fourier domain by 
comparing the power spectrum of the residual data with the input power spectrum 
of the simulated diffuse emission. For discussion on some of the relevant 
methods and an outline of the power spectrum estimation, please see 
\citet{samir14} and references therein.

\begin{figure}
\begin{center}
\psfrag{cl}[b][t][1.5][0]{$C_{\ell} [mK^2]$}
\psfrag{U}[c][c][1.5][0]{$\ell$}
\psfrag{model}[r][r][1][0]{Model}
\psfrag{allnotaper}[r][r][1][0]{Total}
\includegraphics[width=65mm,angle=0]{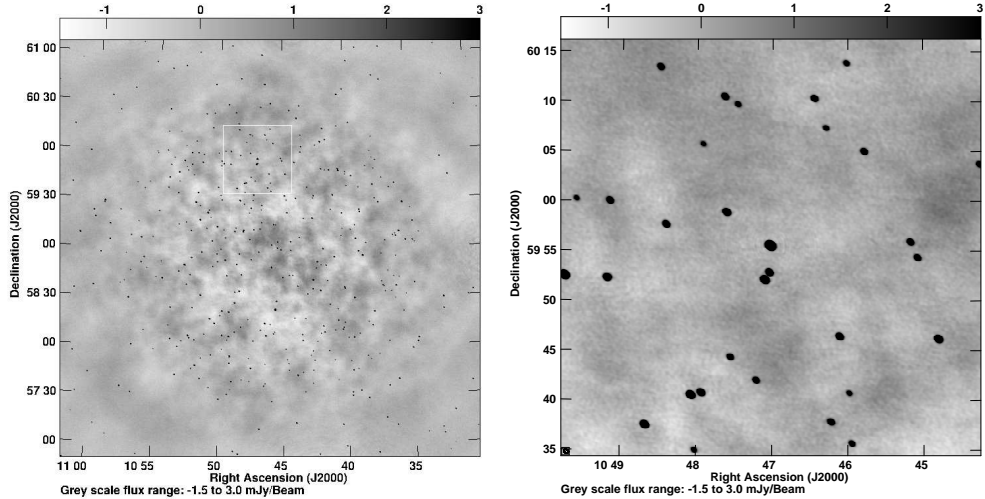}
\includegraphics[width=65mm,angle=0]{img_cleansmall.ps}
\caption{The left panel shows the CLEANed image ($4.2^{\circ}\times4.2^{\circ}$) of the simulated sky centered at R.A.=$10{\rm h} 46{\rm m} 00{\rm s}$ Dec=$59^{\circ} 00^{'} 59^{''}$. The synthesized beam has a ${\rm FWHM} \sim 20^{''}$. A zoom of the square region, $42^{'}\times42^{'}$ in size, marked in the left panel is shown in the right panel. This representative region is used in Figure~\ref{fig:rescompare} for comparison of ``residual'' images. In the central region the  ``off-source'' rms  noise is $\approx 0.3\, {\rm mJy/Beam }$. Here, the grey scale is in units of ${\rm mJy/Beam}$.}
\label{fig:imgclean}
\end{center}
\end{figure}

The left panel of Figure~\ref{fig:imgclean} shows the CLEANed image of the 
simulated sky of the target field with angular size $4.2^{\circ} \times 
4.2^{\circ}$. The synthesized beam has a ${\rm FWHM} \sim 20^{''}$. The image 
contains both point sources and diffuse synchrotron emission, and the grey 
scale flux density range in Figure~\ref{fig:imgclean} is saturated at ${\rm 
3\, mJy}$ to clearly show the diffuse emission. The inner part ($\approx 
1.0^{\circ} \times 1.0^{\circ}$) of CLEANed image has rms noise $\approx 0.3\, 
{\rm mJy/Beam}$, and it drops to $\approx 0.15\,{\rm mJy/Beam }$ at the outer 
part due to the response of the GMRT primary beam attenuation. In the right 
panel of Figure~\ref{fig:imgclean}, we also show a small portion (marked as a 
square box in the left panel) of the image with an angular size $42^{'} \times 
42^{'}$. We note that there is a strong point source at the centre of this 
small image with a flux density of ${\rm 676\, mJy/Beam}$ and spectral index of 
$0.77$. The intensity fluctuations of the diffuse emission are also clearly 
visible in both the panels of Figure~\ref{fig:imgclean}.

\begin{figure}
\begin{center}
\psfrag{cl}[b][t][1.5][0]{$C_{\ell} [mK^2]$}
\psfrag{U}[c][c][1.5][0]{$\ell$}
\psfrag{model}[r][r][1][0]{Model}
\psfrag{allnotaper}[r][r][1][0]{Total}
\includegraphics[width=120mm,height=60mm,angle=0]{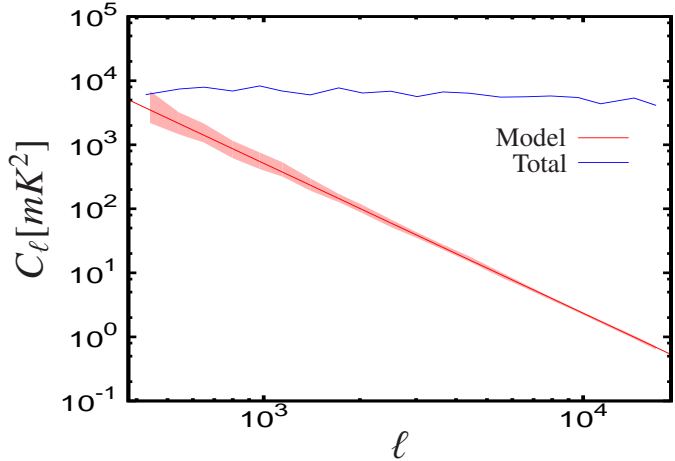}
\caption{The angular power spectrum $C_{\ell}$ estimated from the visibility data which contains both point sources and diffuse synchrotron emission. This spectrum, dominated by the point sources, is flat due to the Poisson distribution of positions of the point sources in our simulation. For comparison, we show the model synchrotron power spectrum (lower curve) with 1-$\sigma$ error estimated from 100 realizations of the diffuse emission map.}
\label{fig:imgcleanps}
\end{center}
\end{figure}

Figure~\ref{fig:imgcleanps} shows the angular power spectrum $C_{\ell}$ 
estimated from the simulated visibilities before any point source subtraction. 
We find that the estimated  power spectrum, as expected, is almost flat across 
all angular scales. This is the Poisson contribution from the randomly 
distributed point sources which dominate $C_{\ell}$ at all angular multipoles 
$\ell$ in our simulation. In this paper, we do not include the clustering 
component of the point sources which becomes dominant only at large angular 
scales ($\ell\le900$) \citep{ali08} where it introduces a power law $\ell$ 
dependence in the angular power spectrum. We also note that the convolution 
with the primary beam affects the estimated angular power spectrum at small 
$\ell$ values (Figure 3, \citealt{samir14}),  and it will be difficult to 
individually distinguish the Poisson and the clustered part of the point source 
components with the GMRT. The total simulated power spectrum $C_{\ell}$ 
(Figure~\ref{fig:imgcleanps}) is consistent with the previous GMRT $150$ MHz 
observations (\citealt{ali08,ghosh150}). In Figure~\ref{fig:imgcleanps} we also 
show the input model angular power spectrum of the diffuse emission along with 
1-$\sigma$ error bar (shaded region) estimated from 100 realizations of the 
diffuse emission map. Note that the angular power spectrum of the diffuse 
emission is buried deep under the point source contribution which dominates at 
all angular scales accessible to the GMRT. We emphasis that, in this paper, our 
aim is to recover this diffuse power spectrum from the residual visibility data 
after point source subtraction.

\subsection{Point Source Subtraction}

\begin{table*}
\begin{center}
\begin{tabular}{|l|c|l|l|}
\hline
{\rm Name}&{\it nterms}&{\rm Threshold flux density}& {\rm CLEANing Box}\\
\hline
{\rm Run(a)}& $1$ &1.0\,{\rm mJy}& Single $4.2^{\circ}\times4.2^{\circ}$ {\rm Box}\\
\hline
{\rm Run(b)}& $2$ &1.0\,{\rm mJy}&Single $4.2^{\circ}\times4.2^{\circ}$ {\rm Box}\\
\hline
{\rm Run(c)}& $2$ &0.5\,{\rm mJy}&Single $4.2^{\circ}\times4.2^{\circ}$ {\rm Box}\\
\hline
{\rm Run(d)}& $2$ &2.0\,{\rm mJy}&Single $4.2^{\circ}\times4.2^{\circ}$ {\rm Box}\\
\hline
{\rm Run(e)}& $2$ &0.5\,{\rm mJy}&Circular region with radius $50^{''}$ \\
& & & around all sources in the image\\
\hline
{\rm Run(f)}& $2$ &2.0\,{\rm mJy}&Single $4.2^{\circ}\times4.2^{\circ}$ {\rm Box}\\
&&0.5\,{\rm mJy}&$1.6^{'}\times1.6^{'}$ {\rm Box} around\\
& & &each visible residual sources\\
\hline
\end{tabular} 
\caption{The set of parameters used for point source imaging with different CLEANing strategies.}
\label{tab:2}
\end{center}
\end{table*}

As shown in Figure~\ref{fig:imgcleanps}, the $150\, {\rm MHz}$ radio sky is 
dominated by point sources at the angular scales $\le 4^{\circ}$ \citep{ali08}. 
Therefore, it is very crucial to identify all point sources precisely from the 
image, and remove their contribution from the visibility data in order to 
estimate the power spectrum of background diffuse emission. However, it is 
quite difficult to model and subtract out the point sources from the sidelobes 
and the outer parts of the main lobe of the primary beam. Our recent paper 
\citep{samir16a} contains a detailed discussion of the real life problems for 
modelling and subtracting point sources from these regions. In this paper we 
have restricted the point source subtraction to the central region of the 
primary beam. To estimate the angular power spectrum $C_{\ell}$ from the 
visibilities, we have used the TGE which tapers the sky response to suppress 
the effect of the point sources outside the FWHM of the primary beam. This is 
achieved by convolving the visibilities with a window function. Note that the 
TGE is also an {\it unbiased} estimator for the angular power spectrum 
$C_{\ell}$; it calculates and subtract the noise bias self-consistently 
\citep[see][for details]{samir14}. Below we discuss the point source modeling 
and the effect of different CLEANing strategies on the ``residual'' images 
created from the point source subtracted visibility data.

\begin{figure*}
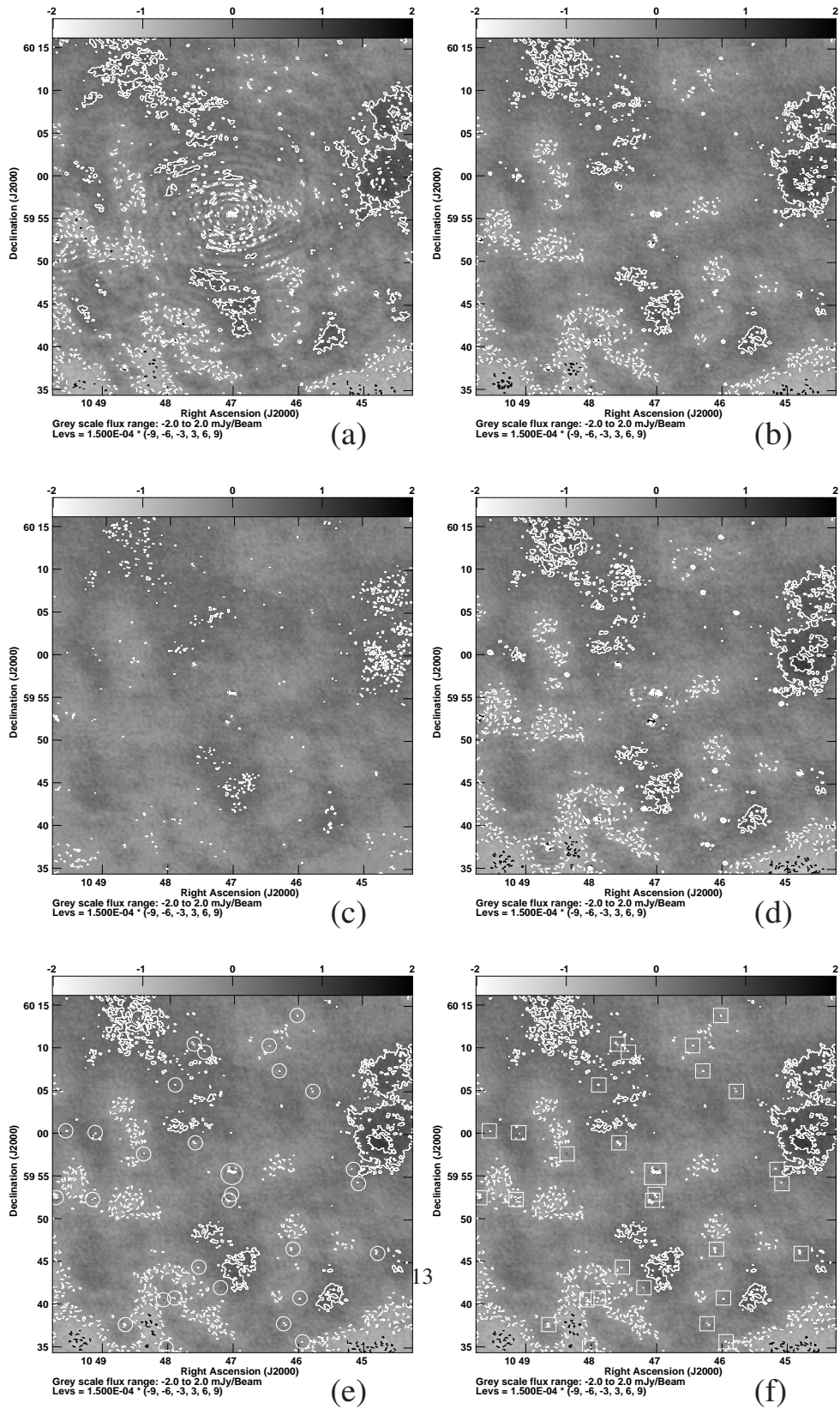

\begin{center}
\includegraphics[width=65mm,angle=0]{img1mjynterm12_2.ps}
\put(-40,5){\Large (a)}
\includegraphics[width=65mm,angle=0]{img1mjy2_2.ps}
\put(-40,5){\Large (b)}

\includegraphics[width=65mm,angle=0]{img0.5mjy2_2.ps}
\put(-40,5){\Large (c)}
\includegraphics[width=65mm,angle=0]{img2mjy2_2.ps}
\put(-40,5){\Large (d)}

\includegraphics[width=65mm,angle=0]{img0.5mjybox2_2star.ps}
\put(-40,5){\Large (e)}
\includegraphics[width=65mm,angle=0]{img2mjyboxhand0.52_2star.ps}
\put(-40,5){\Large (f)}
\caption{Residual images of the $42^{'}\times42^{'}$ representative region for various CLEANing strategies listed in Table~\ref{tab:2}. The residual images Image(a), Image(b), Image(c), ..., and Image(f) correspond to Run(a), Run(b), Run(c), ..., and Run(f) respectively. The contour levels are $(-9,-6,-3,3,6,9)\times0.15{\rm ~mJy/Beam}$ and the grey scale is in ${\rm mJy/Beam}$.}
\label{fig:rescompare}
\end{center}
\end{figure*}

We use standard CASA task CLEAN and UVSUB for deconvolution and removal of 
point sources from the visibility data respectively. CLEAN identifies pixels with flux 
density over the specified threshold, do the deconvolution and create the 
corresponding model visibilities, while UVSUB produce the residual visibility 
by subtracting the model. This should remove the point source contribution from 
the data to a large extent. We further use the residual visibility after point 
source subtraction to make residual ``dirty'' images (without deconvolution) of 
size $4.2^{\circ}\times4.2^{\circ}$. This is done using various CLEANing 
threshold ($0.5$, $1.0$ and $2.0 \,{\rm mJy}$ where $1\, {\rm mJy} \approx 
3\sigma_{im}$), CLEAN box, and spectral modelling options for comparison. For 
CLEAN box, we tried CLEANing the {\it whole} image up to the threshold, or use 
circular boxes of radius $50^{''}$ around {\it all} point sources. As expected, 
the former is more computation expensive and also removes some positive and 
negative peaks of the diffuse signal. On the other hand, the later requires a 
pre-existing deep point source catalogue with accurate position of the sources. 
Note that while such low frequency catalogues for EoR experiments may be 
available from deep continuum surveys in near future, at present it is not a 
realistic strategy. We also used a hybrid method by first CLEANing the whole 
image up to a conservative flux density threshold, and then placing rectangular 
CLEAN boxes of size $1.6^{'}\times 1.6^{'}$ around all residual point sources 
identified visually. These selected regions are then CLEANed up to a deeper flux 
density limit. The effect of spectral modelling is checked by changing the 
parameter ``{\it nterms}'' in the CASA task CLEAN where {\it nterms}=1 does not 
include any spectral correction, while {\it nterms}=2 builds the point source 
model by including spectral index during multi-frequency CLEANing \citep{rau11}.

Table~\ref{tab:2} lists the parameters for a set of CLEANing and point source 
subtraction runs we used for comparison. Figure~\ref{fig:rescompare} shows a 
representative region of angular size $42^{'}\times42^{'}$ from the dirty 
images of the residual data, to illustrate the  effect of different cleaning 
schemes. The different residual images (Image(a) to Image(f)) in 
Figure~\ref{fig:rescompare} correspond to the different CLEANing strategies in 
Table~\ref{tab:2} (Run(a) to Run(f)). The residual images are mostly dominated 
by the diffuse emissions. As expected, correct spectral modelling of the point 
sources significantly improves the residual image as shown clearly in 
Figure~\ref{fig:rescompare} top row (left and right panel for ${\it nterms}=1$ 
and $2$ respectively). Also, CLEANing the whole image to a deeper flux density 
threshold removes part of the diffuse structure. A shallow threshold, on the 
other hand, retains the diffuse emission but also significant residual point 
source contribution (see Figure~\ref{fig:rescompare} middle row). Finally, deep 
CLEANing ($\sim 1.5\,\sigma_{im}$) in combination with carefully selected 
CLEANing regions results in the optimum residual images shown in the bottom 
row of Figure~\ref{fig:rescompare}. In the next section, we assess impact of 
the different CLEAN strategies on the statistics such as distribution of 
visibilities and estimated angular power spectrum from these different residual 
data sets.

\section{Results}
\label{sec:result}

\begin{figure*}
\begin{center}
\psfrag{jy}[c][t][0.8][0]{Flux Density[mJy]}
\psfrag{vjy}[c][t][0.8][0]{Re [Jy]}
\psfrag{number}[b][t][0.8][0]{Number}
\psfrag{tot}[r][r][0.6][0]{Total}
\psfrag{2}[r][r][0.6][0]{{\rm Run(d)}}
\psfrag{1}[r][r][0.6][0]{{\rm Run(b)}}
\psfrag{0.5}[r][r][0.6][0]{{\rm Run(c)}}
\psfrag{0.5bx}[r][r][0.6][0]{{\rm Run(e)}}
\psfrag{0.5bxhnd}[r][r][0.6][0]{{\rm Run(f)}}
\psfrag{gauss}[rb][rb][0.6][0]{Gaussian}
\includegraphics[height=44mm,angle=-90]{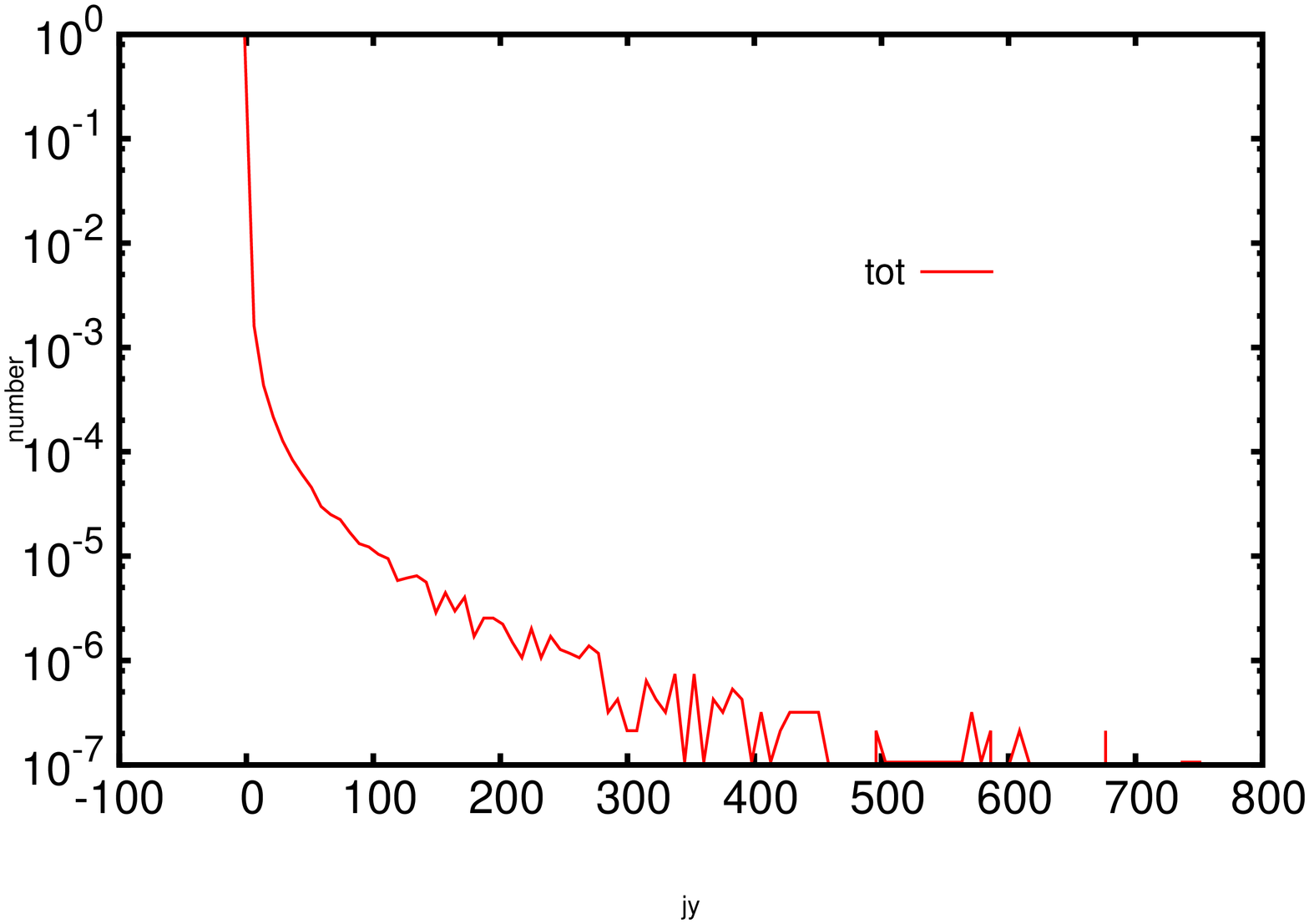}
\includegraphics[height=44mm,angle=-90]{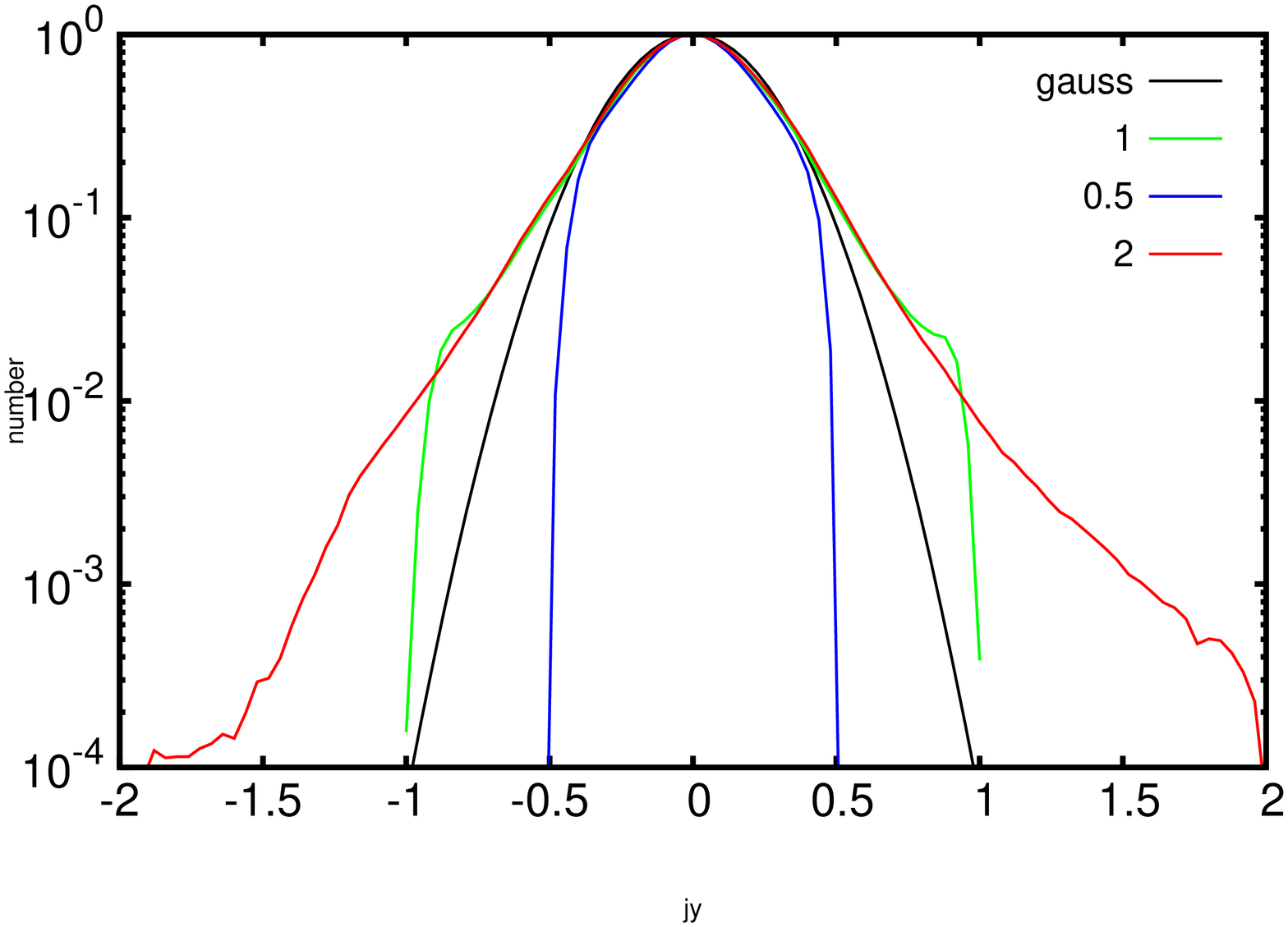}
\includegraphics[height=44mm,angle=-90]{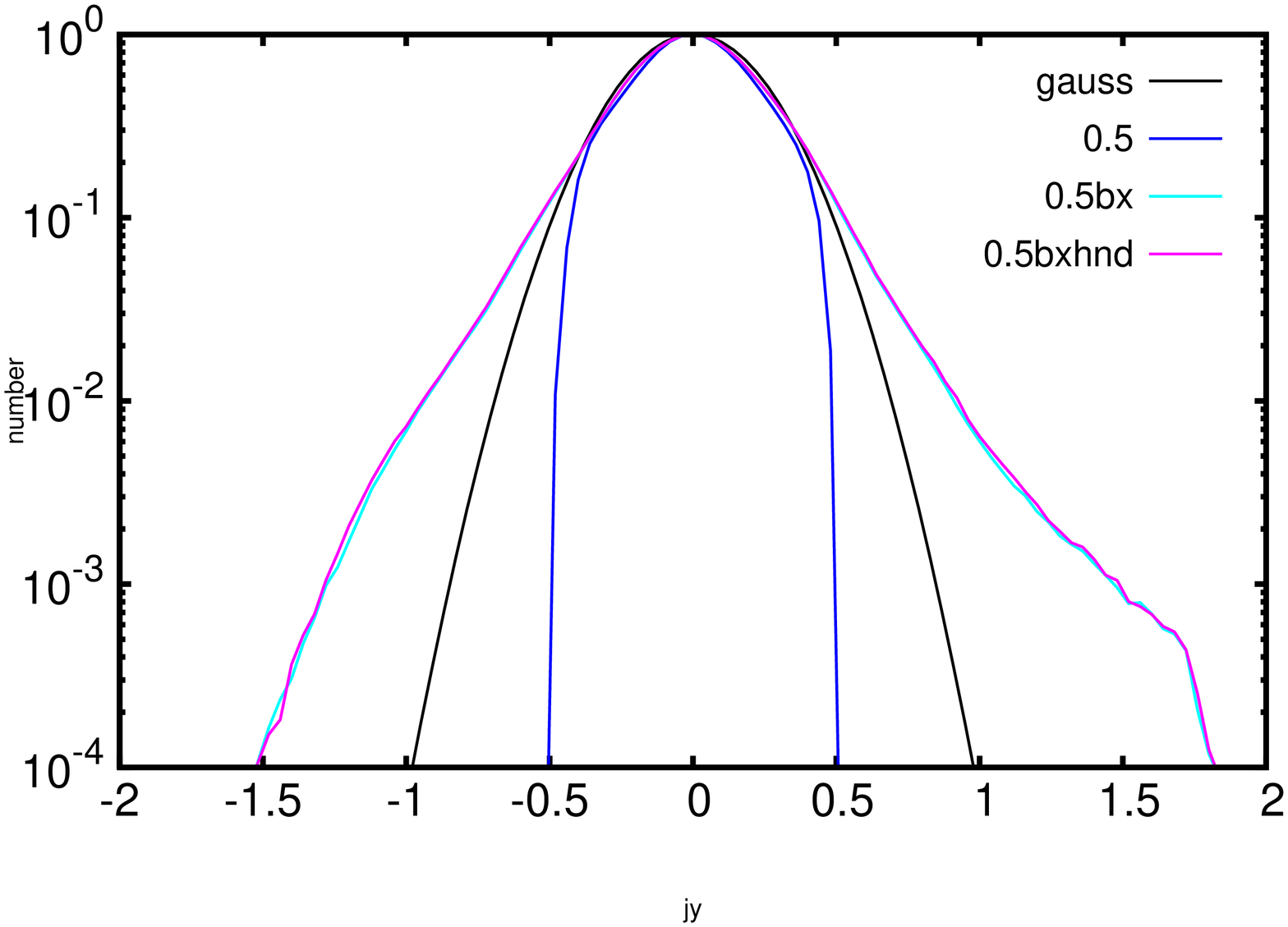}
\includegraphics[height=44mm,angle=-90]{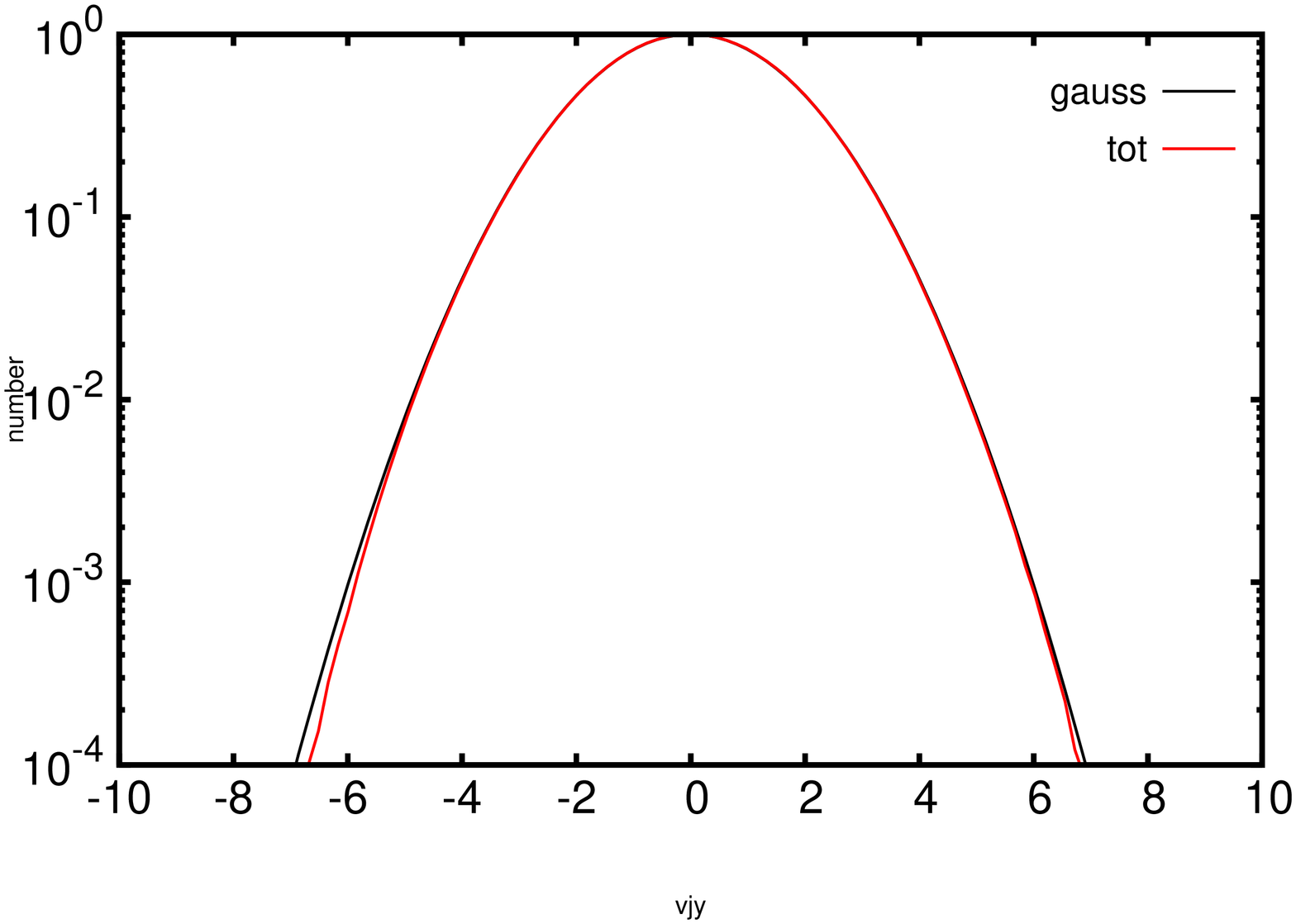}
\includegraphics[height=44mm,angle=-90]{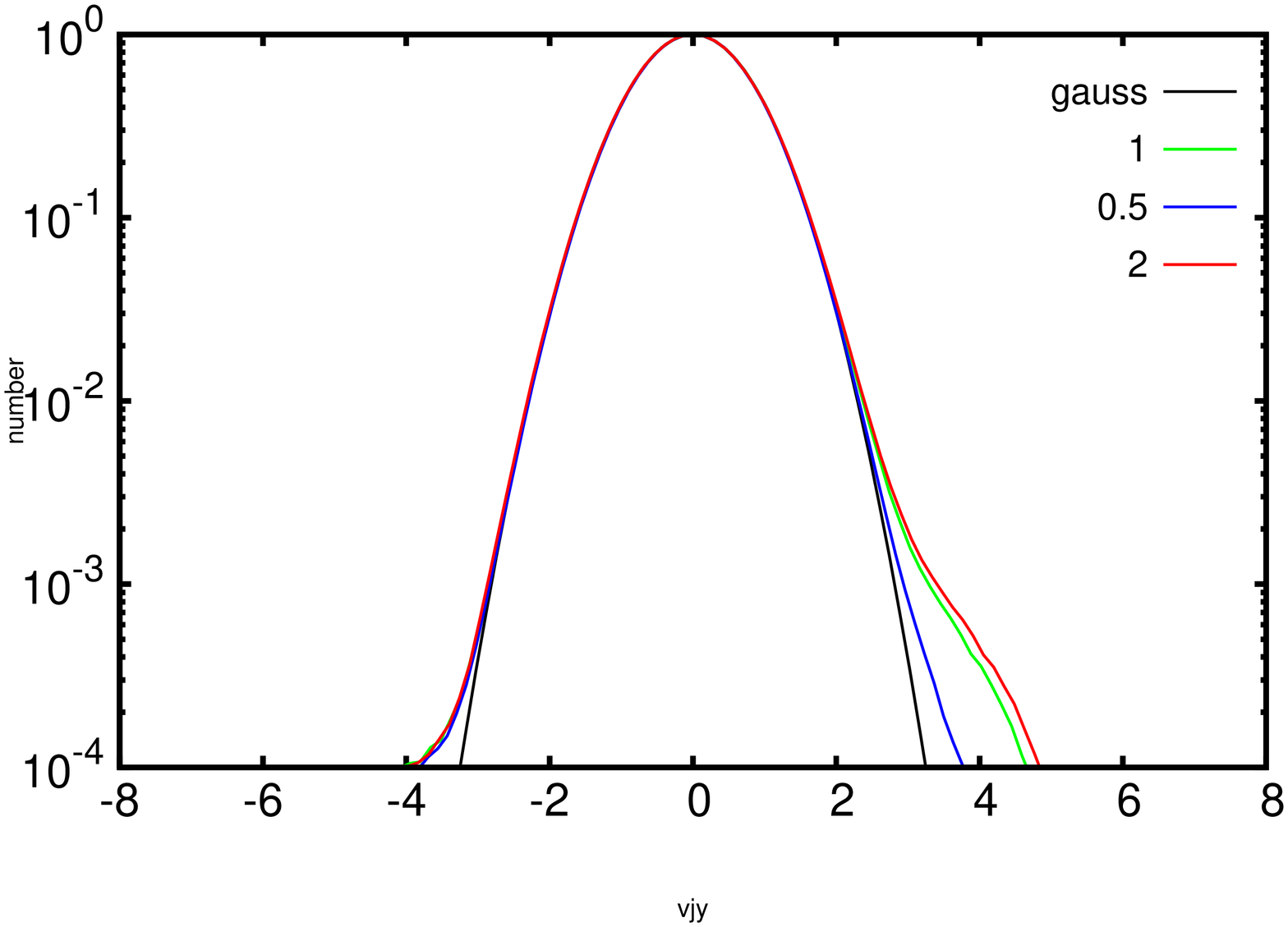}
\includegraphics[height=44mm,angle=-90]{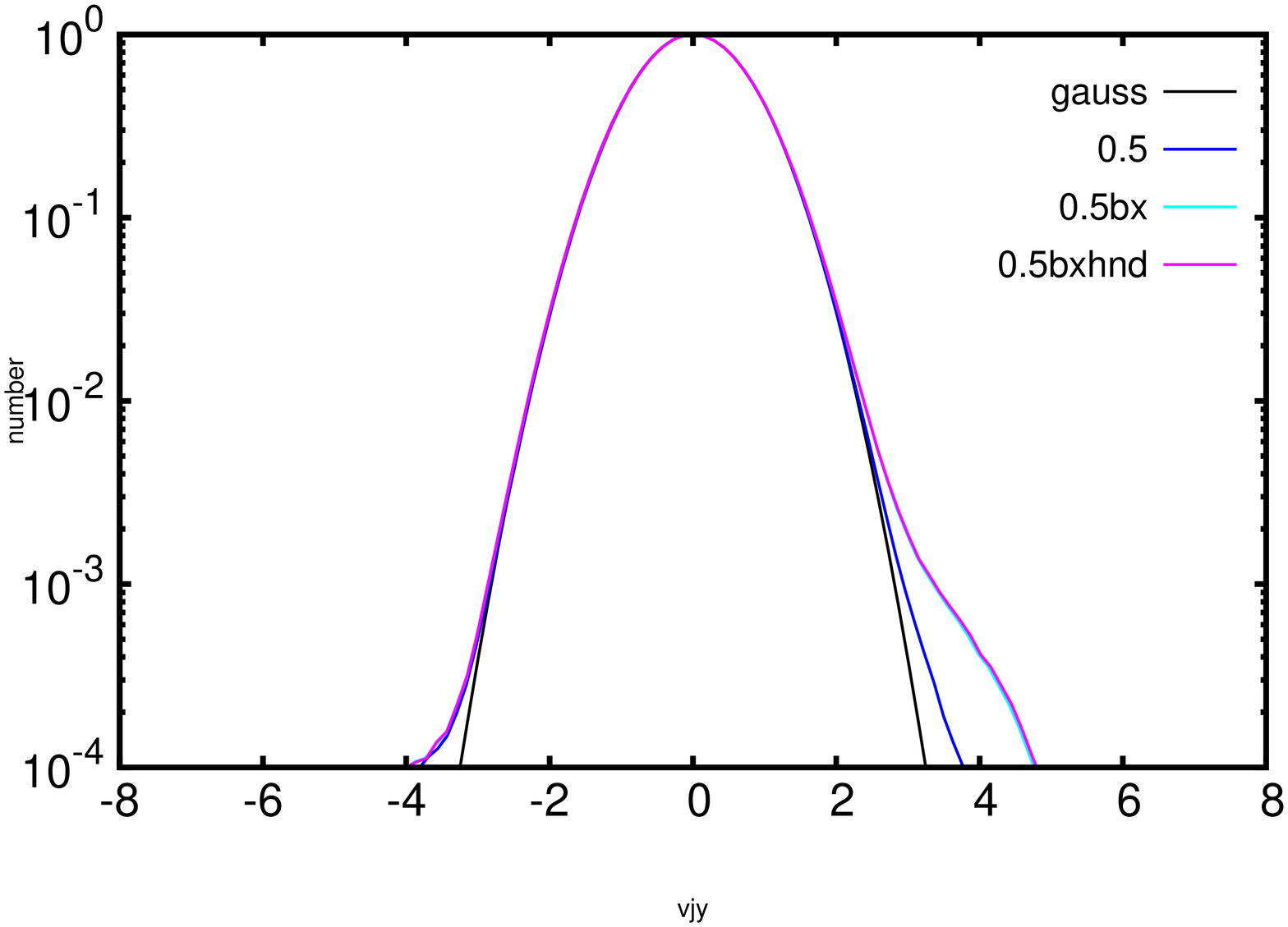}
\caption{The distribution of image plane pixel values (upper row) and the real part of visibilities (lower row) before point source subtraction (left panels) and after point source subtraction (middle and right panels) for the runs mentioned in Table~\ref{tab:2}. The best fit Gaussian function for the distributions are also shown in each panel.}
\label{fig:hist}
\end{center}
\end{figure*}

We use different CLEANing options mentioned above for point source subtraction 
from a $4.2^{\circ} \times 4.2^{\circ}$ region of the sky from simulated 
visibility data. To compare the outcome of these strategies, we check the 
statistics of the residual visibilities as well as of the residual dirty 
images. In Figure~\ref{fig:hist} we show the normalized histograms from images 
(top row) and from the visibility data (bottom row). The top-left panel of 
Figure~\ref{fig:hist} shows the distribution of the pixel values from the 
initial CLEANed map (Figure~\ref{fig:imgclean}) dominated by the diffuse 
emission (pixels with $\le 5.0\, {\rm mJy}$) and only a small number of pixels 
with high flux density (due to the bright point sources). The top middle and 
right panel show the histogram of the residual images from different CLEANing 
runs. A Gaussian with $\sigma=0.228\,{\rm mJy}$ is a fairly good fit to the 
distribution of the residuals up to a flux density limit of $\pm 0.5\, {\rm 
mJy}$. However, as evident from the top central panel, ``blind'' CLEANing with 
lower threshold (see Table~\ref{tab:2}) makes residual images more 
non-Gaussian. On the other hand, for deep CLEANing using different CLEANing box 
options, there is no difference in the distribution of the residual images. 

The corresponding visibility distribution functions are shown in the bottom row 
of Figure~\ref{fig:hist}. We use the real part the complex visibilities for 
these plots, but the imaginary parts also have a similar distribution. We find 
that both the initial and residual visibilities have a Gaussian distribution, 
but with different standard deviation ($\sigma=1.61$ Jy before point source 
subtraction and $0.76$ Jy up to $\mid \rm Re(V)\mid <3{\rm Jy}$ for the residual 
visibility).  The counts significantly deviate from Gaussian distribution at 
large visibility values most likely due to incomplete CLEANing.

\begin{figure}
\begin{center}
\psfrag{cl}[b][t][1.5][0]{$C_{\ell} [mK^2]$}
\psfrag{U}[c][c][1.5][0]{$\ell$}
\psfrag{model}[r][r][1][0]{Model}
\psfrag{nterm1}[r][r][1][0]{{\rm Run(a)}}
\psfrag{nterm2}[r][r][1][0]{{\rm Run(b)}}
\includegraphics[width=80mm,angle=0]{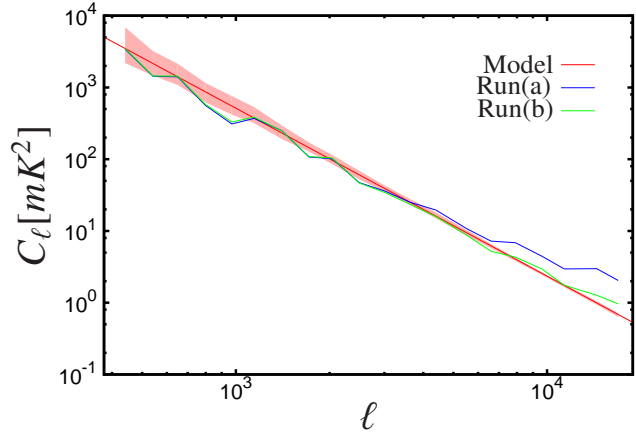}
\caption{The estimated power spectra from residual visibility data for Run(a) and Run(b) corresponding to threshold flux density of $1\,{\rm mJy}$ with ${\it nterms}=1$ and $2$ respectively. The solid line shows the input model (eq. \ref{eq:cl150}) with 1-$\sigma$ error estimated from 100 realizations of the diffuse emission map.}
\label{fig:compnterm}
\end{center}
\end{figure}

Next we use the residual visibilities from different runs to estimate the 
angular power spectrum $C_{\ell}$ using the TGE. Here, we have used logarithmic 
intervals of $\ell$ after averaging all the frequency channels. We have also 
used Gaussian window function to taper the sky response. The tapering is 
introduced through a parameter $f$, where $f$ is preferably $ \le 1$ so that 
modified window function cuts off the sky response well before the first null 
of the primary beam (see for details, Figure 1 of \citealt{samir16a}). The 
reduced field of view results in a larger cosmic variance for the angular modes 
which are within the tapered field of view. So, the tapering parameter $f$ will 
possibly be determined by optimizing between the reduced field of view and the 
cosmic variance. In this work we use $f=0.8$. Figure~\ref{fig:compnterm} shows 
the estimated $C_{\ell}$ from the residual visibilities for Run(a) and Run(b), 
that is for fixed CLEANing threshold of $1.0\, {\rm mJy}$ but ${\it nterms}=1$ 
and $2$ respectively. CLEANing with ${\it nterms}=2$ reduces the residual 
sidelobes around bright sources after point source subtraction (see 
Figure~\ref{fig:rescompare}a,b). Hence, as shown in Figure~\ref{fig:compnterm}, 
the estimated $C_{\ell}$ recover the input power spectrum better even at large 
$\ell$ ($\ge 6\times 10^3$) clearly demonstrating the need of correct spectral 
modelling of the point sources.

\begin{figure}
\begin{center}
\psfrag{cl}[b][t][1.2][0]{$C_{\ell} [mK^2]$}
\psfrag{U}[c][c][1.2][0]{$\ell$}
\psfrag{model}[r][r][0.8][0]{Model}
\psfrag{2mjy}[r][r][0.8][0]{{\rm Run(d)}}
\psfrag{1mjy}[r][r][0.8][0]{{\rm Run(b)}}
\psfrag{0.5mjy}[r][r][0.8][0]{{\rm Run(c)}}
\psfrag{o.5mjybox}[r][r][0.8][0]{{\rm Run(e)}}
\psfrag{0.5mjyboxhand}[r][r][0.8][0]{{\rm Run(f)}}
\includegraphics[width=67mm,angle=0]{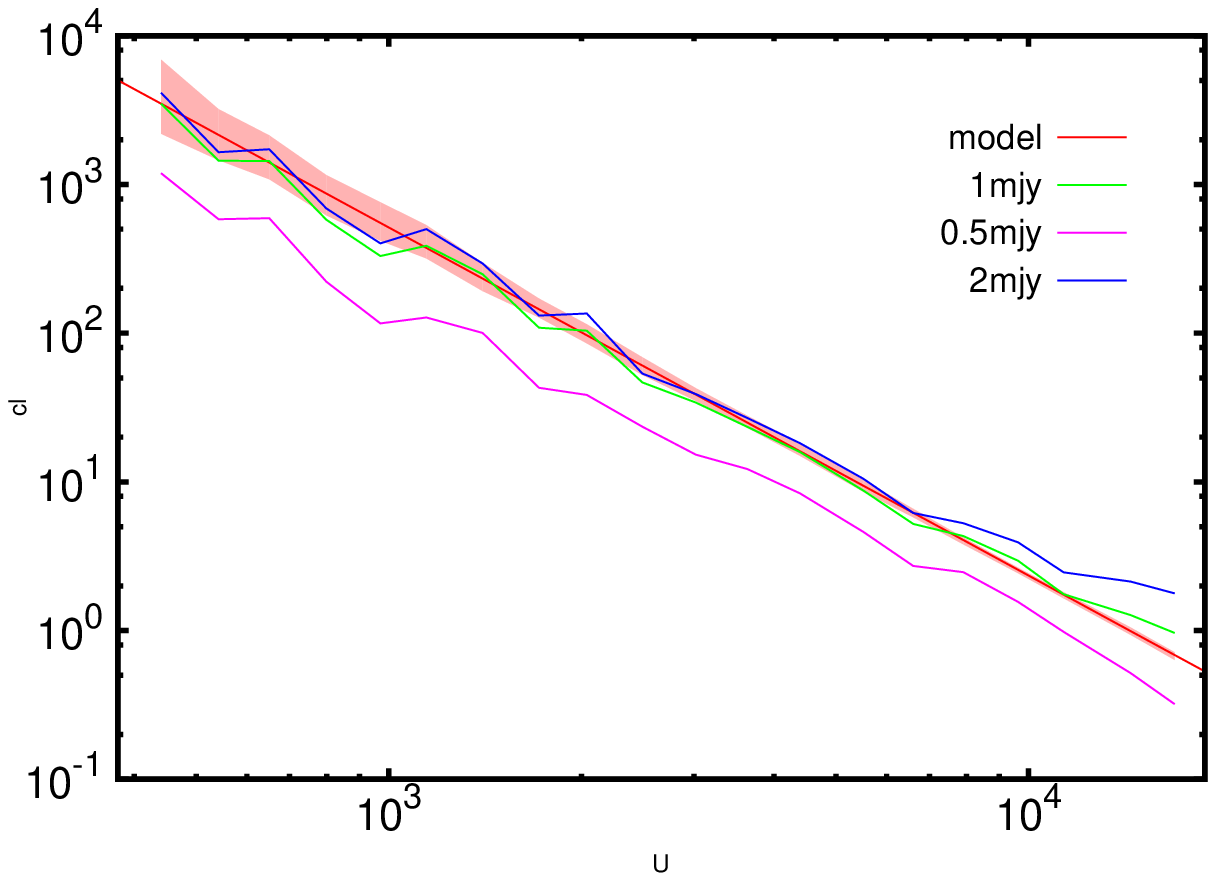}
\includegraphics[width=67mm,angle=0]{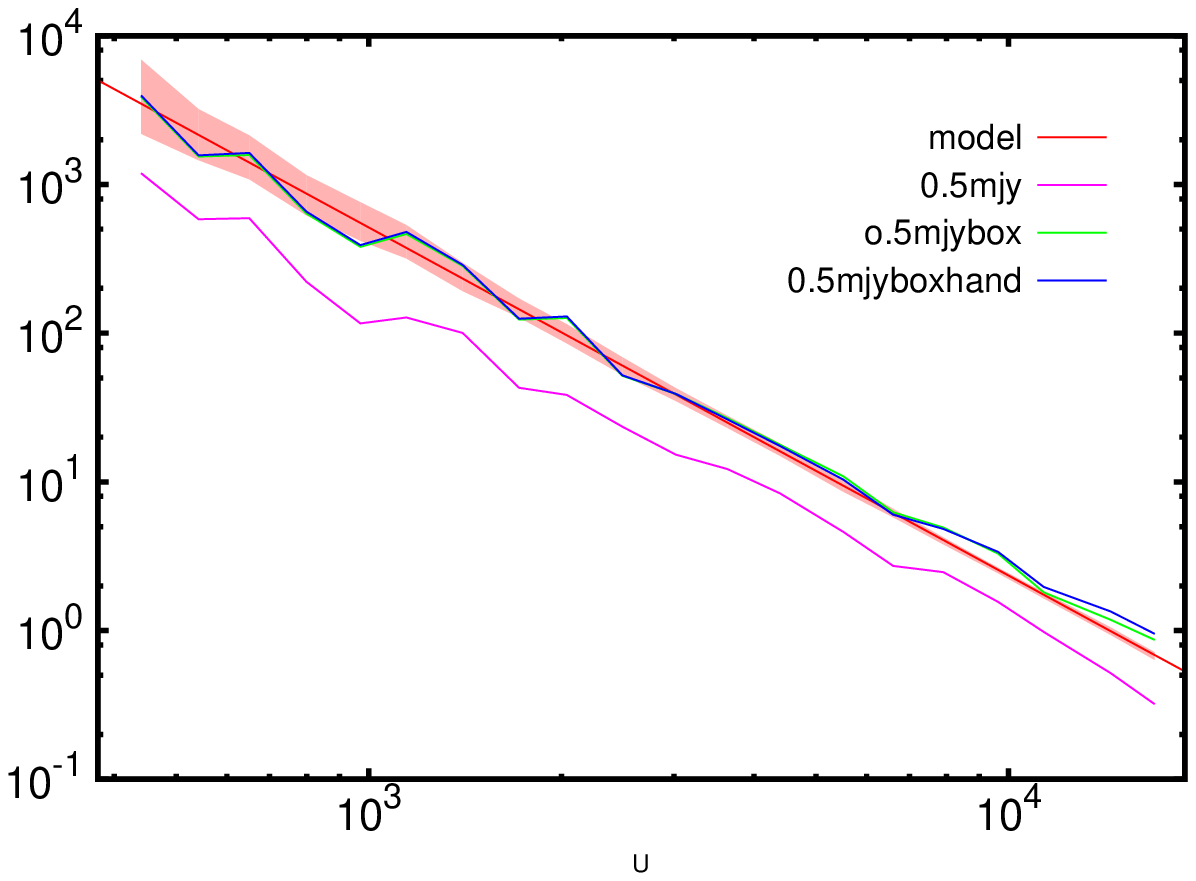}
\caption{The left panel show the estimated power spectra using residual data from Run(b), (c) and (d) with different CLEANing threshold but fixed value of ${\it nterms}=2$. The right panel show the same but using residual data from Run(c), (e) and (f) with different CLEAN box options but fixed CLEANing threshold (see Table \ref{tab:2} for details).}
\label{fig:compcutoff}
\end{center}
\end{figure}

The left panel of Figure~\ref{fig:compcutoff} shows the angular power spectra $C_{\ell}$ 
estimated using the residual visibility data obtained from Run(b), (c) and (d) 
for ${\it nterms}=2$ but different CLEANing threshold. Run(b) with $\sim 3\, 
\sigma_{im}$ CLEANing threshold recovers $C_{\ell}$ for the entire range of 
$\ell$, but Run(d) with shallow CLEANing retains some extra residual power at 
large $\ell$ ($\ge 7\times 10^3$). The estimated $C_{\ell}$ from Run(c), on the 
other hand, falls off by a factor $\sim 5$ compared to the input model due to 
blind deep CLEANing that removes part of the underlying diffuse signal. The 
effect of using different CLEANing box options in recovering $C_{\ell}$ is 
shown in the right panel of Figure~\ref{fig:compcutoff}. Here we keep the other two parameters fixed 
at ${\it nterms}=2$ and threshold of $0.5\,{\rm mJy}$. It is clear from this 
figure that there is no significant change in the estimated power spectra for 
the two different CLEANing box strategies used in Run(e) and (f). In both the 
cases, the estimated $C_{\ell}$ agree well with the input power spectrum over 
the full range of $\ell$ probed here.

\section{Summary and conclusions}
\label{summa}

Precise subtraction of point sources from wide-field interferometric data is 
one of the primary challenges in studying the diffuse foreground emission as 
well as the weak redshifted HI 21-cm signal. In this paper, we demonstrate the 
method of studying and characterizing the Galactic synchrotron emission using 
simulated $150\, {\rm MHz}$ GMRT observation in presence of point sources. The 
angular power spectrum $C_{\ell}$ of the diffuse emission is estimated from the 
residual visibility data using TGE after subtracting the point sources from 
only the inner part of the field of view. The estimated $C_{\ell}$ due to faint point sources is much lower than the diffuse synchrotron emission. We assess the impact of imperfect 
point source removal for different CLEANing strategies in recovering the input 
$C_{\ell}$ of the diffuse emission for the angular scale range probed by the 
GMRT.

The simulations are carried out for GMRT $150 \, {\rm MHz}$ observation for a 
sky model consisting of point sources and diffuse synchrotron emission. The sky 
model is multiplied with the model PB ${\cal A}(\th,\,\nu)$, before computing 
the visibilities for the frequency and the $uv$ coverage of the simulated GMRT 
observation. We use various CLEANing strategies with different CLEANing boxes, 
threshold flux and spectral correction options to make images and to subtract 
point source model from the simulated visibilities. The residual data were then 
used for estimating $C_{\ell}$ of the diffuse component. We check the effect of 
point source subtraction by comparing image histograms, visibility distribution 
function as well as $C_{\ell}$ from the residual data. 

We find that all the different CLEANing strategies introduce some degree of 
non-Gaussianity in the residual data both in image and in visibility domain. 
The less precise point source subtraction generates more non-Gaussianity in the 
distribution of image-pixels beyond the CLEANing threshold. Equivalently, the 
visibility distributions also deviate significantly from a Gaussian. Comparing 
the recovered and the input power spectra, we find that both shallow CLEANing 
and incorrect spectral modelling of the point sources result in excess power at 
the large angular multipoles. On the other hand, very deep ``blind'' CLEANing 
removes part of the diffuse structure and reduces the amplitude of the power 
spectrum at all angular scale. Carefully choosing CLEAN boxes for deep CLEANing 
(with threshold $\sim 1.5\sigma_{im}$) and correct spectral modelling of the 
point sources demonstrate that TGE can recover the input power spectrum of the 
diffuse emission properly. Note that this analysis also demonstrate that the 
effect of the point sources from the outer region of the field is insignificant 
due to the tapering. Hence, while using TGE for power spectrum estimation, many 
of the complications discussed earlier related to the low frequency wide field 
imaging become irrelevant. 

Finally, the accurate removal of all the point sources from the wide-field 
image is complicated and difficult task in presence of instrumental 
systematics, calibration errors, RFI and ionospheric effects etc. Using 
simulated data, we have established here the effectiveness of TGE in estimating 
the angular power spectrum $C_{\ell}$ of diffuse emission at the angular scales 
probed by the GMRT. This gives us the confidence to apply it on real data in 
order to study the Galactic synchrotron power spectrum \citep{samir16c}. With 
the broad goal of applying it in future for EoR and post-EoR HI studies, we 
plan to next incorporate some of the above mentioned ``real world'' issues in 
this simulation, and also extend this study for the SKA. 

\section{Acknowledgements}

SC would like to acknowledge the University Grant Commission (UGC), India for 
providing financial support through Senior Research Fellowship. SSA would like 
to acknowledge CTS, IIT Kharagpur for the use of its facilities. SSA would also 
like to thank the authorities of the IUCAA, Pune, India for providing the 
Visiting Associateship programme. AG would like acknowledge Postdoctoral Fellowship from the South African Square Kilometre Array Project for financial support. {\bf PD will like to acknowledge the DST-INSPIRE faculty 
fellowship by Department of Science and Technology, India for providing financial support.}

\bibliographystyle{model2-names}
\bibliography{references}
\end{document}